\documentclass[12pt,epsf,axodraw]{article}

\newcommand{\bea}{\begin{eqnarray}}
\newcommand{\eea}{\end{eqnarray}}
\newcommand{\be}{\begin{equation}}
\newcommand{\ee}{\begin{equation}}

\begin{document}

\thispagestyle{empty}

\begin{center}
{\LARGE \bf Single and double see-saw for 
quark-lepton and neutrino masses\\}
\vspace{2cm}
{\bf Biswajoy Brahmachari}
\\
\vskip 1cm
{
Department of Physics, Vidyasagar Evening College\\
39, Sankar Ghosh Lane, Kolkata 700006, India.\\}
\end{center}
\vskip 1cm
\begin{abstract}
A left-right model of quarks and leptons based on the gauge group 
$SU(3)_C \times SU(2)_L \times SU(2)_R \times U(1)_{B-L}$ is studied.
Here the scalar sector consists of only two doublets namely 
(1,2,1,1) and (1,1,2,1) but familiar bidoublet (1,2,2,0) is removed. 
Quarks and charged leptons get mass from a single see-saw mechanism but   
neutrinos get mass by a double see-saw mechanism. In this type of models
the heaviest right handed neutrino can be of the order 
of $10^{13}$ GeVs or less in a natural way, depending on the size 
of related Yukawa couplings.
\end{abstract}
\vskip 2.5in
\begin{flushright}
VEC/PHYSICS/P/1/2003-2004
\end{flushright}

\newpage

When the particle content of the standard model is extended by including 
right handed gauge bosons \cite{lr}, the gauge group becomes ${\cal G}_{LR}
\equiv SU(3)_C \times SU(2)_L \times SU(2)_R \times U(1)_{B-L}$, and a
linear combination of diagonal generators lead to conserved electric
charge at low energy. The relationship is
\begin{equation}
Q = T_{3L} + T_{3R} + \frac{(B-L)}{2} = T_{3L} + \frac{Y}{2}. \label{eq3}
\end{equation}
Quarks and leptons transform under $G_{LR}$. It is not always easy to
embed low energy quarks and leptons in representations of larger gauge
groups. But for the left-right symmetric case, $G_{LR}$ posesses natural
representations to embed the quarks and leptons of the standard model.
\begin{eqnarray}
q_L = (u,d)_L &\sim& (3,2,1,1/3), \label{eq04} \\ 
q_R = (u,d)_R &\sim& (3,1,2,1/3), \label{eq4} \\ 
l_L = (\nu,e)_L &\sim& (1,2,1,-1), \label{eq05} \\ 
l_R = (N,e)_R &\sim& (1,1,2,-1), \label{eq5}
\end{eqnarray}
where a new fermion, i.e.~$N_R$, has been added to fill
the gap in $l_R$ doublet. This is just the right handed neutrino. For three
generations, there are three of them. The lightest mass eigenstate of
this right handed triplet may decay violating CP (Charge conjugation
times Parity), to produce a tiny lepton asymmetry in early 
universe\cite{lepto}.

In previous left-right models\footnote{
See Ref\cite{bm89} for models without (1,2,2,0)}, 
a scalar bidoublet transforming as 
$(1,2,2,0)$ was included because then we get correct Yukawa couplings
at  the tree level leading to fermion masses. It is a natural choice 
if we want the quarks and leptons to have tree-level masses. But fermion 
mass problem seems to be a much more involved one. Fermions may 
not have got masses at the tree level at all. Instead they may have got 
masses from higher dimensional operators. The details of such high scale
physics could be so obscure to us at the moment, 
that one may have to resort to effective non-renormalizable operators 
of mass dimension more than four.  

Suppose however that we are only interested in 
the spontaneous breaking of $SU(2)_L \times SU(2)_R \times U(1)_{B-L}$ to 
$U(1)_{em}$ with $v_R >> v_L$, then the simplest way is to introduce two 
Higgs doublets transforming as
\begin{eqnarray}
\Phi_L = (\phi^+_L,\phi^0_L) &\sim& (1,2,1,1), \label{eq06} \\ 
\Phi_R = (\phi^+_R,\phi^0_R) &\sim& (1,1,2,1). \label{eq6}
\end{eqnarray}
Suppose we now do not admit any other scalar multiplet
\footnote{This Higgs choice was made previously by
Babu and Mohapatra\cite{bm89} in the context of the
strong CP problem}. This situation can be compared with 
to the situation in the standard model, where $SU(2)_L \times U(1)_Y$ is 
spontaneously broken down to $U(1)_{em}$ by a Higgs doublet and we do not 
admit anything else in the scalar sector.  In that case, we 
find that quark and  charged-lepton masses are generated at tree level, 
but neutrinos obtain Majorana masses only through the dimension-five 
operator\cite{seesaw}. In our case, 
in the absence of the bidoublet, all fermion masses,  
have their origin in some kind of see-saw mechanism\footnote{A summary
of various mechanisms to obtain neutrino mass can be found in
Ref. \cite{maprl}}, as shown below. Using Eqs.~(\ref{eq04}) 
to (\ref{eq6}), it is clear that the following two objects
\begin{eqnarray}
(l_L \Phi_L) &=& \nu_L \phi_L^0 - e_L \phi_L^+ \label{eq7}, \\
(l_R \Phi_R) &=& N_R \phi_R^0 - e_R \phi_R^+  \label{eq8},
\end{eqnarray}
are invariants under ${\cal G}_{LR}$.  Hence we have the dimension-five 
operators given by
\begin{equation}
{\cal L}_M = {f^L_{ij} \over 2 \Lambda_M} (l_{iL} \Phi_L)(l_{jL} \Phi_L) 
+ {f^R_{ij} \over 2 \Lambda_M} (l_{iR} \Phi_R)(l_{jR} \Phi_R) + H.c.,
\label{eq9}
\end{equation}
which will generate Majorana neutrino masses proportional to $v_L^2/\Lambda_M$ 
for $\nu_L$ and $v_R^2/\Lambda_M$ for $N_R$. In addition, we have
\begin{equation}
{\cal L}_D = {f^D_{ij} \over \Lambda_D} (\bar l_{iL} \Phi_L^*)
(l_{jR} \Phi_R) + H.c.,
\label{eq10}
\end{equation}
and the corresponding dimension-five operators which will generate Dirac 
masses for all the quarks and charged leptons. Therefore the combination
$(\Phi^*_L \Phi_R)$ behaves as an effective (1,2,2,0) scalar \cite{prl}
eventhough we have forbidden (1,2,2,0) in our Higgs choice.

Because in left-right symmetric models, both left handed as well as
right handed projections of neutrino is available at our
disposal, they can pair up and, from Eq.~(\ref{eq10}) we get the 
Dirac type mass of neutrino as,
\begin{equation}
(m_D)_{ij} = {f^D_{ij} v_L v_R \over \Lambda_D}, \label{eq11}
\end{equation}
hence $\nu_L$ gets a double seesaw Majorana \cite{dss} mass of order
\begin{equation}
{m_D^2 \over m_N} 
\sim {v_L^2 v_R^2 \over \Lambda_D^2} {\Lambda_M \over v_R^2} 
= {v_L^2 \Lambda_M \over \Lambda_D^2}, \label{eq12}
\end{equation}
where in Eq. (\ref{eq12}) we have used the usual single seesaw formula
\begin{equation}
M_{light}=M_L+m^T_{Dirac}{1 \over M_R} m_{Dirac}.
\end{equation}
Note that ${m_D^2 \over m_N}$ is much larger than 
$v_L^2/\Lambda_M \equiv M_L$ if $\Lambda_D << \Lambda_M$.  
Take for example $\Lambda_M$ to be the Planck scale of $10^{19}$ GeV and 
$\Lambda_D$ to be the grand-unification scale of $10^{16}$ GeV, then the 
neutrino mass scale is 1 eV for $v_L$ of order 100 GeV. We can also
see that $M_L \sim 10^{-6}$ eV can be safely neglected.  
These masses are tabulated in Table \ref{table1}. The difference 
between $\Lambda_M$ and $\Lambda_D$ may be due to the fact that if we assign 
a global fermion number $F$ to $l_L$ and $l_R$, then ${\cal L}_M$ has $F = 
\pm 2$ but ${\cal L}_D$ has $F=0$.

Since the Dirac masses of quarks and charged leptons are also given by 
Eq. (\ref{eq11}), $v_R$ cannot be much below $\Lambda_D$.  This means 
that $SU(2)_R \times U(1)_{B-L}$ is broken at a very high scale 
to $U(1)_Y$, and our model at low energy is just the standard model.  
We do however have the extra right handed neutrinos $N_R$ with masses of 
order $v_R^2/\Lambda_M$, i.e. below $10^{13}$ GeV, which are useful for 
leptogenesis, as is well-known.

\begin{table}
\begin{center}
       \begin{tabular}{|c|c|c|c|}
      \hline
      $v_L$/GeV & $\Lambda_M$/GeV & $\Lambda_D$/GeV & $m_\nu$/eV \\
       \hline
       100 & $10^{19}$ & $10^{16}$ & 1 \\
       100 & $10^{18}$ & $10^{16}$ & 0.1 \\
       100 & $10^{18}$ & $2 \times 10^{16}$ & 0.025 \\
        91 & $10^{18}$ &  $2 \times 10^{16}$ & 0.0207 \\
       \hline
       \end{tabular}
\end{center}
\caption{Magnitudes of neutrino masses generated via 
double seesaw}
\label{table1}
\end{table}

\section*{Acknowledgements}
I would like to thank Prof. R.N. Mohapatra for communications

\end{document}